\documentclass[prl,twocolumn,superscriptaddress,secnumarabic,balancelastpage,amsmath,amssymb,nofootinbib]{revtex4-1}
\usepackage{amssymb}
\usepackage{amsmath}
\usepackage{lgrind}        
\usepackage{color}         
\usepackage{graphics}      
\usepackage[pdftex]{graphicx}      
\usepackage{longtable}     
\usepackage{epsf}          
\usepackage{bm}            
\usepackage{asymptote}     
\usepackage{thumbpdf}
\usepackage{verbatim}
\usepackage{url}
\usepackage[colorlinks=true]{hyperref}  
\usepackage{enumitem}	

\usepackage{float}
\usepackage{times}
\usepackage{color}
\usepackage{verbatim}

\begin{document}

\title{Observation of Bulk Fermi Arc and Polarization\\ Half Charge from Paired Exceptional Points}

\author{Hengyun Zhou}
\thanks{These authors contributed equally to this work.}
\affiliation{Department of Physics, Massachusetts Institute of Technology, Cambridge, Massachusetts 02139, USA}
\affiliation{Department of Physics, Harvard University, Cambridge, Massachusetts 02138, USA}
\author{Chao Peng}
\thanks{These authors contributed equally to this work.}
\affiliation{Department of Physics, Massachusetts Institute of Technology, Cambridge, Massachusetts 02139, USA}
\affiliation{State Key Laboratory of Advanced Optical Communication Systems \& Networks, Peking University, Beijing 100871, China}
\author{Yoseob Yoon}
\affiliation{Department of Chemistry, Massachusetts Institute of Technology, Cambridge, Massachusetts 02139, USA}
\author{Chia Wei Hsu}
\affiliation{Department of Applied Physics, Yale University, New Haven, Connecticut 06520, USA}
\author{Keith A. Nelson}
\affiliation{Department of Chemistry, Massachusetts Institute of Technology, Cambridge, Massachusetts 02139, USA}
\author{Liang Fu}
\affiliation{Department of Physics, Massachusetts Institute of Technology, Cambridge, Massachusetts 02139, USA}
\author{John D. Joannopoulos}
\affiliation{Department of Physics, Massachusetts Institute of Technology, Cambridge, Massachusetts 02139, USA}
\author{Marin Solja\v{c}i\'{c}}
\affiliation{Department of Physics, Massachusetts Institute of Technology, Cambridge, Massachusetts 02139, USA}
\author{Bo Zhen}
\thanks{To whom correspondence should be addressed; E-mail: bozhen@sas.upenn.edu.}
\affiliation{Department of Physics, Massachusetts Institute of Technology, Cambridge, Massachusetts 02139, USA}
\affiliation{Department of Physics and Astronomy, University of Pennsylvania, Philadelphia, Pennsylvania 19104, USA}

\begin{abstract}
The ideas of topology have found tremendous success in Hermitian physical systems, but even richer properties exist in the more general non-Hermitian framework. Here, we theoretically propose and experimentally demonstrate a new topologically-protected bulk Fermi arc which---unlike the well-known surface Fermi arcs arising from Weyl points in Hermitian systems---develops from non-Hermitian radiative losses in photonic crystal slabs. Moreover, we discover half-integer topological charges in the polarization of far-field radiation around the Fermi arc. We show that both phenomena are direct consequences of the non-Hermitian topological properties of exceptional points, where resonances coincide in their frequencies and linewidths. Our work connects the fields of topological photonics, non-Hermitian physics and singular optics, and paves the way for future exploration of non-Hermitian topological systems.
\end{abstract}


\maketitle

In recent years, topological physics has been widely explored in closed and lossless Hermitian systems, revealing novel phenomena such as topologically non-trivial band structures \cite{CastroNeto2009,Armitage2017,Tarruell2012,Wan2011,Xu2015,Lv2015,Lu2015,Noh2017,Yang2017} and promising applications including back-scattering-immune transport \cite{Hasan2010,Qi2011,Lu2014,Haldane2008,Wang2009,Fang2012,Hafezi2013,Rechtsman2013a,Khanikaev2013,Yang2015,Susstrunk2015,Wu2015}. 
However, most systems, particularly in photonics, are generically non-Hermitian due to radiation into open space or material gain/loss. 
Non-Hermiticity enables even richer topological properties, often with no counterpart in Hermitian frameworks \cite{Moiseyev2011,Cao2015,Leykam2017,Shen2017}.
One such example is the emergence of a new class of degeneracies, commonly referred to as exceptional points (EPs), where two or more resonances of a system coalesce in both eigenvalues and eigenfunctions \cite{Berry2004a,Rotter2009,Heiss2012}.
So far, isolated EPs in parameter space \cite{Dembowski2001,Mailybaev2005,Liertzer2012,Gao2015,Doppler2016,Xu2016,Hassan2017} and continuous rings of EPs in momentum space \cite{Zhen2015,Cerjan2016a,Xu2017} have been studied across different wave systems due to their intriguing properties, such as unconventional transmission/reflection \cite{Lin2011a,Feng2013,Longhi2014}, relations to parity-time symmetry \cite{Bender1998,Makris2008,Guo2009,Ruter2010,Chong2011,Regensburger2012,Konotop2016}, as well as their unique applications in sensing \cite{Hodaei2017,Chen2017} and single-mode lasing \cite{Hodaei2014,Feng2014,Peng2014a}.

Here, we theoretically design and experimentally realize a new configuration of isolated EP pairs in momentum space, which allows us to reveal the unique topological signatures of EPs in the band structure and far-field polarization, and to extend topological band theory into the realm of non-Hermitian systems.
Specifically, we demonstrate that a Dirac point (DP) with nontrivial Berry phase can split into a pair of EPs \cite{Kirillov2005,Seyranian2005,Kozii2017} when radiation loss---a form of non-Hermiticity---is added to a 2D-periodic photonic crystal (PhC) structure. The EP-pair generates a distinct double-Riemann-sheet topology in the complex band structure, which leads to two novel consequences: bulk Fermi arcs and polarization half charges.
First, we discover and experimentally demonstrate that this pair of EPs is connected by an open-ended isofrequency contour---we refer to it as a bulk Fermi arc---in direct contrast to the common intuition that isofrequency contours are necessarily closed trajectories.
The bulk Fermi arc here is a unique topological signature of non-Hermitian effects in paired EPs, and resides in the bulk dispersion of a 2D system.
This is fundamentally different from the previously known surface Fermi arcs that arise from the 2D projection of Weyl points in 3D Hermitian systems.
Moreover, we find experimentally that near the Fermi arc frequency, the system exhibits a robust half-integer winding number in the far-field polarization \cite{Zhan2009,Berry2003,Freund2005,Miyai2006,Dennis2009,Bauer2015,Fosel2017}, analogous to the orientation-reversal on a Mobius strip.
We show that this is a direct consequence of the topological band-switching properties across the Fermi arc connecting the EP pair, and is direct experimental evidence of the $\nu=\pm1/2$ topological index associated with an EP \cite{Shen2017}.
With comprehensive comparisons between analytical models, numerical simulations, and experimental measurements, our results are a direct validation of non-Hermitian topological band theory, and present its novel application to the field of singular optics.

\begin{figure*}[htb]
\begin{center}
\includegraphics[width=12cm]{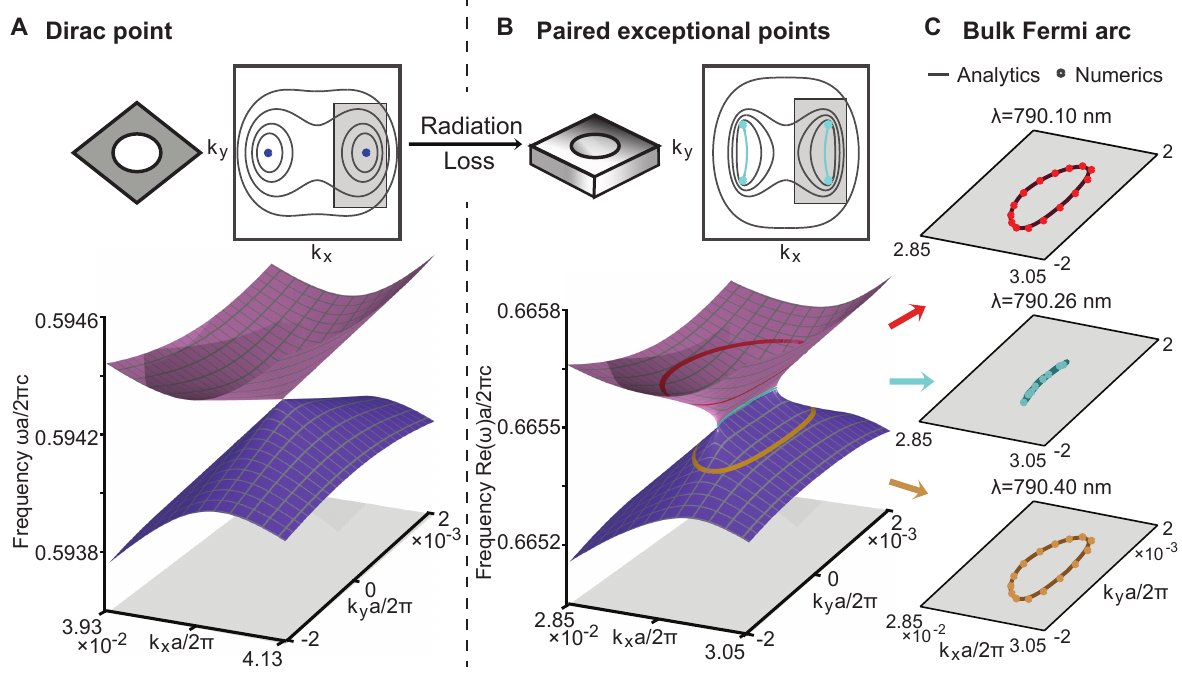}
\end{center}
\caption{{\bf Bulk Fermi arc arising from paired exceptional points split from a single Dirac point.}
{\bf A-B,} Illustration of photonic crystal (PhC) structures, isofrequency contours, and band structures;
{\bf A,} Band structure of a 2D-periodic PhC consisting of a rhombic lattice of elliptical air holes, featuring a single Dirac point on the positive $k_x$ axis.
{\bf B,} The real part of the eigenvalues of an open system consisting of a 2D-periodic PhC slab with finite thickness, where resonances experience radiation loss. The Dirac point splits into a pair of exceptional points (EPs). The real part of the eigenvalues are degenerate along an open-ended contour---the bulk Fermi arc (blue line)---connecting the pair of EPs.
{\bf C,} A few examples of the isofrequency contours in this system, including the open bulk Fermi arc at the EP frequency (middle panel), and closed contours at higher (upper panel) or lower (lower panel) frequencies. Solid lines are from the analytical model, and circles are from numerical simulations.}
\end{figure*}

We start by outlining our scheme to split a single DP into a pair of EPs, which directly leads to the emergence of a bulk Fermi arc. First, consider a 2D-periodic photonic crystal (PhC) with a square lattice of circular air holes introduced into a dielectric material.
In this Hermitian system (no material gain/loss or radiation loss), the crystalline symmetry ($C_{4v}$) ensures a quadratic band degeneracy at the center of the Brillouin zone (Supplementary Fig.~S1A). As this $C_{4v}$ symmetry is broken, e.g. by shearing the structure into a rhombic lattice with elliptical holes (Fig.~1A), the quadratic degeneracy point splits into a pair of DPs situated at $(\pm k_\text{D}, 0)$ along the $k_{x}$ axis. The same splitting behavior is shown in both analytical models and numerical simulations (Supplementary Information, Section I) \cite{Chong2008,Zhou2016} .

Next, we consider a non-Hermitian system of a finite thickness PhC slab (inset of Fig.~1B), where modes near the DP become resonances with finite lifetime due to radiative losses towards the top and bottom.
Adopting the even and odd $y$-mirror-symmetric eigenstates at the DP as basis, and taking into account the radiation losses via non-Hermitian perturbations, the effective Hamiltonian in the vicinity of the original DP at $(k_\text{D}, 0)$ can be written as \cite{Zhou2016,Shen2017}:  
\begin{equation} 
\label{eqn:NH-Hamiltonian}
H_{\rm eff} = \omega_\text{D} - i\gamma_0 + (v_g \delta k_x- i\gamma) \sigma_z + v_g \delta k_y \sigma_x,
\end{equation} 
with complex eigenvalues of 
\begin{equation} 
\label{eqn:EPring}
\omega_{\pm} = \omega_\text{D} - i\gamma_0 \pm \sqrt{(v_g^2 \delta k^2 -\gamma^2) - 2i\gamma v_g\delta k_x}.
\end{equation} 
Here $\sigma_{x,z}$ are Pauli matrices, $\omega_{\text{D}}$ is the DP frequency, and $(\delta k_{x}, \delta k_{y})$ is the momentum displacement from $(k_\text{D},0)$, $\delta k^2 = \delta k_x^2 + \delta k_y^2$. Meanwhile, $\gamma_{0} \pm \gamma$ are the radiation decay rates of even and odd $y$-mirror-symmetric modes, taking into account the fact that the two modes have different coupling strengths to the continuum; $v_g$ is the group velocity describing the dispersion around the DP, which for simplicity is here chosen to be the same along all directions (see Supplementary Information, Sections I \& II, for the general case). The real part of the complex eigenvalues $\omega_{\pm}$ characterizes the resonance frequency, while the imaginary part represents the linewidth of the resonance.

The eigenvalue spectrum exhibits a pair of EPs at $(\delta k_{x},\delta k_{y})= (0, \pm \gamma/v_{g})$ when the square root term in Eq.~(2) vanishes and the two eigenvectors coalesce (Fig.~1B). Furthermore, this pair of EPs are connected in momentum space by an open-ended arc---a bulk Fermi arc, along which the real part of the complex eigenvalues are degenerate at $\omega_\text{D}$ (Fig.~1C middle panel).
Although sharing features similar to previously-studied Fermi arcs---both are open-ended isofrequency contours---our bulk Fermi arc resides in the bulk dispersion rather than on the surface of a 3D Hermitian Weyl system, and originates from non-Hermiticity rather than the presence of Weyl points.
As the frequency $\omega$ decreases from above $\omega_\text{D}$, the closed isofrequency contour at $\omega$ shrinks (Fig.~1C top panel), eventually turning into the open Fermi arc when $\omega = \omega_\text{D}$ (Fig.~1C middle panel), and expands out again into a closed contour at even lower frequencies (Fig.~1C bottom panel).
Taken together, the band structure around the EPs forms a double-Riemann-sheet topology (Fig.~1B). This originates from the complex square root term in the dispersion in Eq.~(2), which, depending on the sign choice of the square root, results in two sheets. The two eigenvalues continuously evolve on each sheet, and their real parts become degenerate along a curve---the bulk Fermi arc.
We further verify the existence of bulk Fermi arcs in realistic PhC slab structures via numerical simulations (Fig.~1C circles), showing a good agreement with analytical results (Fig.~1C solid lines); see Supplementary Information, Sections II \& III, for details.

\begin{figure}[htb]
\begin{center}
\includegraphics[width=6cm]{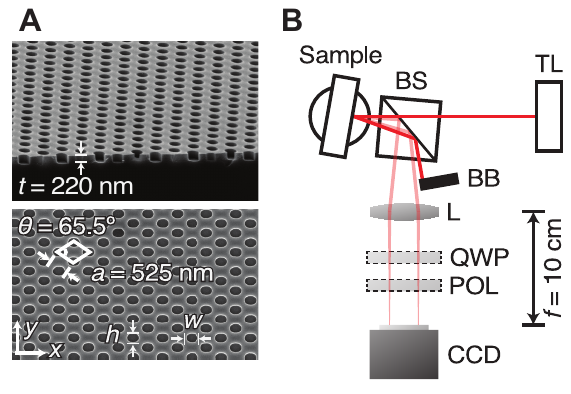}
\end{center}
\caption{{\bf Fabricated PhC slab and measurement setup.} 
{\bf A,} SEM images of the PhC samples: side view (top panel) and top view (bottom panel).
{\bf B,} Schematic drawing of the scattering measurement setup. Vertically-polarized light from a tunable continuous-wave Ti-Sapphire laser (TL) scatters off the PhC slab and is collected using a CCD camera placed at the focal plane of the lens (L). The specular reflection is blocked to ensure only scattered light is imaged. POL and QWP are used in polarimetry measurements of the scattered light. 
TL, tunable laser; BS, beam-splitter; L, convex lens with $10$ cm focal length; POL, polarizer; QWP, quarter wave plate; BB, beam block.} 
\end{figure}

To experimentally demonstrate the bulk Fermi arc, we use interference lithography to fabricate PhC slabs in Si$_3$N$_4$ (refractive index $n = 2.02$, thickness $t=220$ nm) on top of a silica substrate ($n = 1.46$) \cite{Lee2012}. The PhC structure consists of rhombic unit cells with side length $a=525$ nm, unit cell angle $\theta=65.5^\circ$, and elliptical air holes with long axis length $w=348$ nm and short axis length $h=257$ nm (see Supplementary Information, Section IV, for details). Scanning electron microscope (SEM) images of the fabricated samples are shown in Fig.~2A. The structure is immersed in an optical liquid with refractive index matched to that of the silica substrate to create an up-down symmetric environment.  

We performed angle-resolved scattering measurements (setup shown in Fig.~2B) to image isofrequency contours of the sample. 
The PhC sample is illuminated with a tunable continuous-wave Ti:Sapphire laser (MSquared) that is vertically polarized, while scattered light---arising from natural fabrication imperfections of the sample---is collected with a CCD camera (Princeton Instruments) placed at the focal plane of a convex lens with $10$ cm focal length. 
Due to resonant-enhancement, the scattered light will have strongest intensity only along directions where the underlying resonances share the same frequency as the pump laser, and thus the isofrequency contours of the sample are directly imaged onto the CCD \cite{Regan2016,Shi2010} (see Supplementary Information, section IV, for details of the setup and the scattering process).
To show the full shrinking-and-reexpanding feature of the isofrequency contours around the bulk Fermi arc, the laser wavelength is tuned from $794$ nm down to $788$ nm at steps of approximately $0.2$ nm.
Furthermore, the polarization at each point along a given isofrequency contour is determined through polarimetry measurements, by optionally inserting a quarter-wave plate and a polarizer (Thorlabs) in front of the CCD (see Supplementary Information, Section V, for details). 

 \begin{figure}[htb]
\begin{center}
\includegraphics[width=8cm]{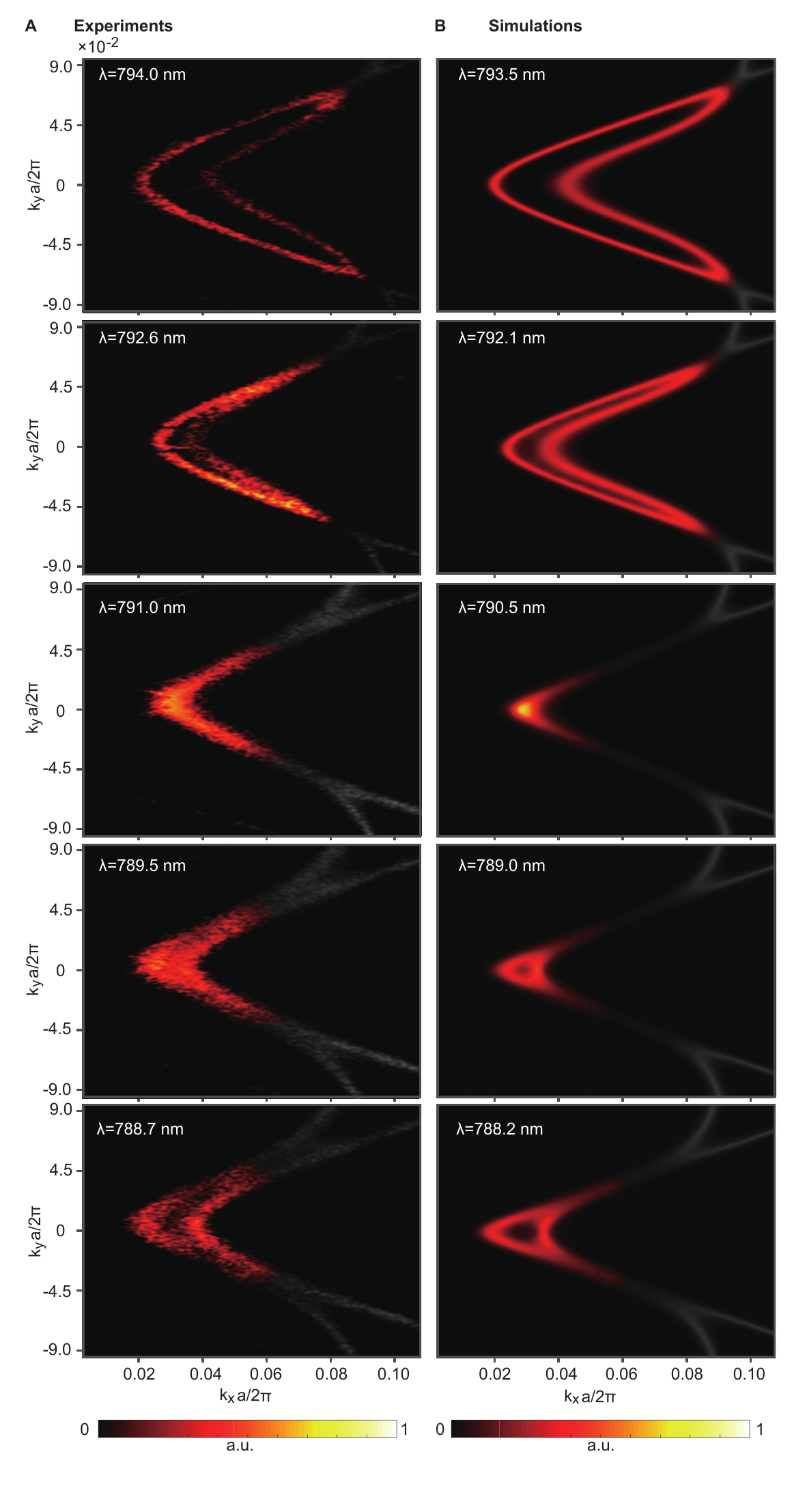}
\end{center}
\caption{{\bf Experimental demonstration of a bulk Fermi arc.} 
{\bf A,} Experimentally measured isofrequency contours and {\bf B,} numerically simulated spectral density of states at five representative wavelengths. The bulk Fermi arc appears at $791.0$ nm (middle row), when the isofrequency contour becomes open-ended. The regions of interest are highlighted in all panels to emphasize the shrinking (top two rows) and re-expanding (bottom two rows) feature of isofrequency contours near the bulk Fermi arc. The numerical results are offset by $0.5$ nm for better comparison.
}
\end{figure}

In Fig.~3A, the experimental results of isofrequency contours are shown at a few representative wavelengths around the Fermi arc. These are compared to numerical results of isofrequency contours (Fig.~3B) obtained from simulating fitted structural parameters, showing a good agreement with each other.
Here, for better comparison, the numerical results are offset by $0.5$ nm relative to the experiments. 
See Supplementary Information, section IV, about this wavelength offset, and Supplementary Movie I for the full set of isofrequency contours measured at different wavelengths. 
To focus on the bulk Fermi arc, we highlight the region of interest in both panels, where the isofrequency contours clearly demonstrate the shrinking and re-expanding behavior.
As shown in Fig.~3, as the wavelength decreases from $794.0$ nm, the corresponding isofrequency contour shrinks (top two rows), and eventually becomes an open-ended arc at $791.0$ nm (middle row), consistent with our previous theoretical predictions in Fig.~1C. As the wavelength is further decreased down to $789.5$ nm and $788.7$ nm, the arc expands out into closed contours again (bottom two rows). 
The bending feature of the contours is a result of higher-order terms in the band dispersion (Supplementary Information section II).
The open contour at $791$ nm (middle row) is a clear, direct observation of the bulk Fermi arc.

So far, we have shown one direct consequence of the unique double-Riemann-sheet topology near paired EPs---the bulk Fermi arc. Next, we demonstrate another consequence: half-integer topological charges in the polarization configuration, which also serve as a direct experimental validation of the $\nu = \pm 1/2$ topological index of an EP. These topological charges describe the direction (clockwise or counterclockwise) and number of times the polarization vector winds around a point/line singularity in the optical field, and in our particular system we observe a robust 180-degree winding around the Fermi arc, corresponding to a half-integer charge.

To fully reconstruct the far-field polarization configurations of the resonances, we perform polarimetry measurements by recording the intensity of isofrequency contours after passing through six different configurations of polarizers and/or waveplates (see Supplementary Information, section V, for details). Although the incoming light is vertically polarized, the scattered light at each point along the contour is, in general, elliptically polarized, reflecting the polarization state of its underlying resonance \cite{Hsu2017}. Taking points X and Z in Fig.~4A as examples: after passing through a vertical polarizer, the scattered light is weak(strong) at point X(Z); while after a horizontal polarizer, the relative intensity of the scattered light switches between points X and Z. This clearly shows that the far field of the underlying resonance at point X(Z) is mostly horizontally(vertically) polarized.

Examples of the fully-reconstructed spatial polarizations (blue ellipses) at representative points along the $794$ nm isofrequency contour (red solid line) are shown in the top panel of Fig.~4B, which agree well with numerical results (Fig.~4B bottom panel). Furthermore, both experimental and numerical results show 180-degree winding of the polarization long axis, as illustrated by the green arrows in Fig.~4B: as the momentum point starts from point X, traverses the full contour in the counterclockwise direction, and returns to point X, the polarization long axis flips direction by rotating $180$ degrees in the clockwise direction---corresponding to a $-1/2$ topological charge being enclosed in the loop. These results thus indicate that the far-field emission from our PhC is a vector-vortex beam with half-integer topological charge, in stark contrast to the integer vector beams realized in photonic crystal surface emitting lasers \cite{Miyai2006,Iwahashi2011}.

\begin{figure*}[htb]
\begin{center}
\includegraphics[width=12cm]{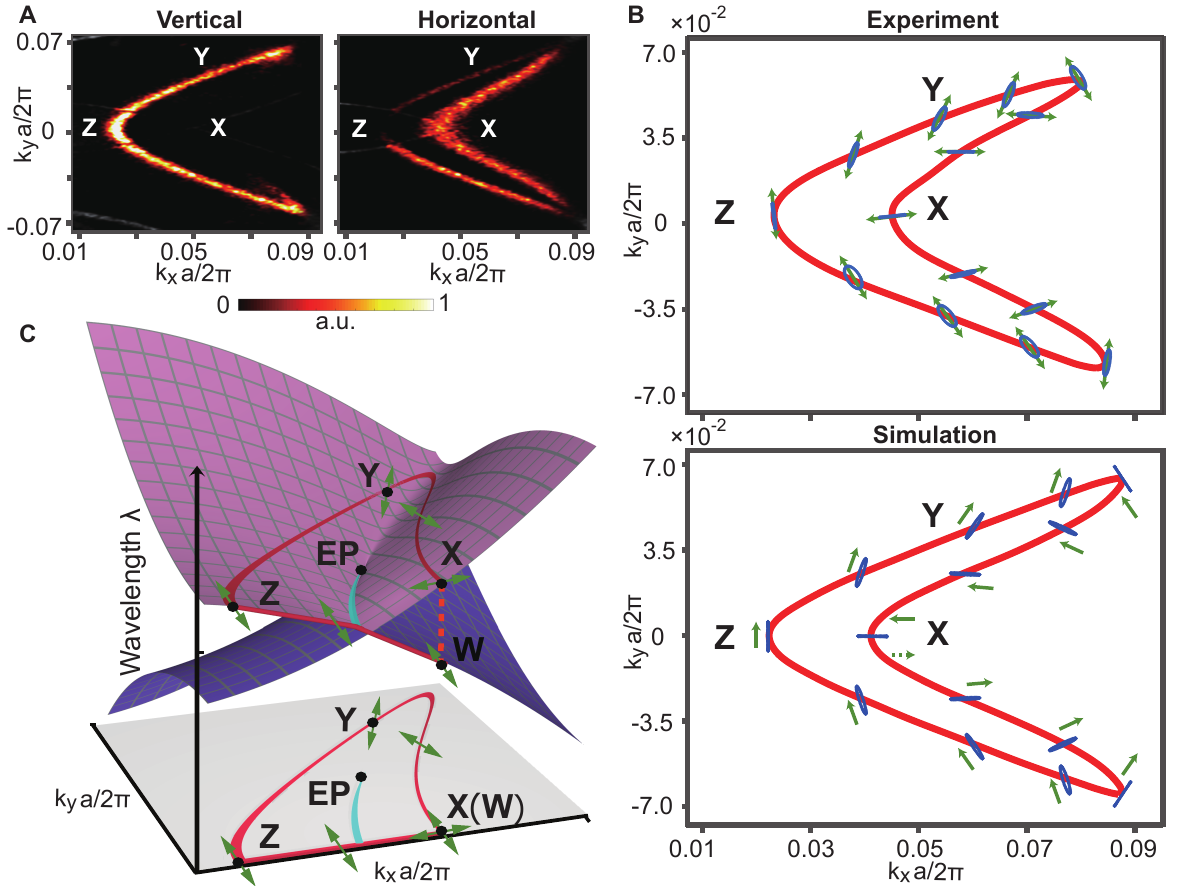}
\end{center}
\caption{{\bf Experimental demonstration of polarization topological half charges around the bulk Fermi arc.} 
{\bf A,} Intensity of scattered light at 794 nm after passing through a vertical (horizontal) polarizer is plotted in the left (right) panel, showing that point X(Z) is mostly horizontally (vertically) polarized.
{\bf B,} Experimental reconstruction (top) and numerical simulation (bottom) of the full polarization information, showing the polarization ellipses (blue ellipses) as well as their long axis directions (green arrows) along an isofrequency contour (red line). As shown by the green arrows in the bottom panel, the polarization long axis exhibits a $-1/2$ topological charge.
{\bf C,} Schematic illustration of the mode switching (X to W) in the band structure, along a loop enclosing an EP (X-Y-Z-W), as a result of the double-Riemann-sheet topology. This mode switching behavior directly leads to the half-integer topological index of an EP and the half-charge polarization winding.
}
\end{figure*}

We now elaborate on the fundamental connections between the half-integer topological charges observed in the far-field polarization and the half-integer topological index of an EP \cite{Shen2017}, manifested as its mode-switching property. For more details see Supplementary Information, section VI. 
Along the $k_x$ axis, the two bands forming the EP pair in our system have orthogonal linear polarizations due to the $y$-mirror symmetry: one is horizontal (e.g. mode X in Fig.~4C), while the other is vertical (e.g. mode Z and W).
As we follow a closed path in momentum space X$\rightarrow$Y$\rightarrow$Z$\rightarrow$W that encircles one of the EPs in the counterclockwise direction, the initial eigenstate X (horizontally polarized) on the top sheet adiabatically evolves into state Z (vertically polarized) and eventually into final state W (vertically polarized) on the bottom sheet, due to the mode-switching topological property of the EP \cite{Dembowski2001,Doppler2016,Xu2016,Hassan2017}.
The switching behavior of the eigenmodes---from X to W---directly follows from their eigenvalue swapping behavior on the complex plane (see Supplementary Section II and Fig.~S2).
Equivalently, one complex eigenvalue winds around the other one by half a circle, thus implying that the topological index of an EP is a half-integer.
The orthogonal nature between the polarizations at X 
and Z, arising from the mode-switching property of the EP, guarantees a $(n+1/2)\pi$-rotation ($n\in \mathbb{Z}$) of the polarization vector along half the contour. Again using the $y$-mirror symmetry, the full isofrequency contour will accumulate twice the rotation angle, to a combined $(n+1/2)\times 2\pi$-rotation, corresponding to a half-integer charge of $n+1/2$.

We have thus shown the intimate connection between polarization vector winding in singular optics and the double-Riemann-sheet topology of paired EPs. Our experimental demonstration---generating half-integer vector-vortex beams directly from the topological properties of EPs---not only distinguishes us from the previously known integer topological charges of polarization around bound states in the continuum \cite{Zhen2014}, but also proves the nontrivial topology of EPs.

In this work, we have demonstrated that the topological properties of paired EPs endow the band structure and far-field emission with unique features, manifested as the emergence of bulk Fermi arcs and polarization half charges.
Application-wise, our structure provides a simple-to-realize method to create half-integer vector-vortex beams \cite{Dorn2003,Zhan2009} at a wide range of frequencies.
Future prospects leveraging the topological landscape around paired EPs may enable PhC lasers with exotic emission profiles \cite{Cai2012,Hirose2014}, such as twisted Mobius strips.
The isolated EPs found in our structure also provide a straightforward platform to study the influence of EPs and their topology on light-matter interactions, such as modified Purcell factors for spontaneous emission enhancement and nonlinear optics generation \cite{Lin2016,Pick2017,Pick2017a}.
Finally, our study paves the way for the exploration of topological band theory in general non-Hermitian wave systems, ranging from photonic and acoustic to electronic and polaritonic systems.

\section*{Acknowledgments}
The authors thank Dr. Tim Savas for fabrication of the samples. The authors also thank Steven Johnson, Eugene Mele, Ling Lu, Yichen Shen, Scott Skirlo, Shang Liu, Jong Yeon Lee, Francisco Machado, Nicholas Rivera, Grace Zhang and Sam Moore for helpful discussions. Research supported as part of the Army Research Office through the Institute for Soldier Nanotechnologies under contract no. W911NF-13-D-0001 (photon management for developing nuclear-TPV and fuel-TPV mm-scale-systems). Also supported as part of the S3TEC, an Energy Frontier Research Center funded by the US Department of Energy under grant no. DE-SC0001299 (for fundamental photon transport related to solar TPVs and solar-TEs). H.Z. acknowledges support from the Undergraduate Research Opportunities Program at MIT. C.P. acknowledges support from the National Natural Science Foundation of China under Grants No. 61575002, 61320106001, and the China Scholarship Council. C.W.H. acknowledges support from the National Science Foundation under grant NSF DMR-1307632. Y.Y. and K.A.N. were supported in part by Skoltech as part of the Skoltech Next Generation Program.

\section*{Author Contributions}
H.Z. and B.Z. conceived the idea. H.Z. performed the analytical calculations and numerical simulations. H.Z., B.Z., C.P. and Y.Y. conducted the experiments and analyzed the data. H.Z. and B.Z. wrote the manuscript, with input from all authors. B.Z., M.S. and J.D.J. supervised the research. All authors contributed to the analysis and discussion of the results.

\end{document}